\documentclass[12pt,letterpaper]{article}

\usepackage{amsmath,amssymb,calc}
\usepackage{graphicx}
\usepackage{cite}

\newcommand\be{\begin{equation}}
\newcommand\ee{\end{equation}}
\newcommand{\bea}{\begin{eqnarray}}
\newcommand{\eea}{\end{eqnarray}}

\newcommand{\nn}{\nonumber}
\newcommand{\pd}{\partial}

\def\id{\protect{{1 \kern-.28em {\rm l}}}}

\def\id{\protect{{1 \kern-.28em {\rm l}}}}

\begin{document}

\begin{titlepage}
\begin{center}
\hfill \\
\vspace{2cm}
{\Large {\bf On the Consistency of Rapid-Turn Inflation\\[3mm] }}

\vskip 1.5cm
{\bf Lilia Anguelova,\\
\vskip 0.5cm  {\it Institute for Nuclear Research and Nuclear Energy,\\
Bulgarian Academy of Sciences, Sofia 1784, Bulgaria}\\
{\tt anguelova@inrne.bas.bg}}

\vskip 6mm

\end{center}

\vskip .1in
\vspace{1cm}

\begin{center} {\bf Abstract}\end{center}

\vspace{-1cm}

\begin{quotation}\noindent

Recent studies, in the context of consistency conditions for rapid-turn and third order slow-roll inflation in two-field models, raised the question whether this regime can be sustained for more than a few e-folds of expansion. We answer this question in the affirmative by showing that the consistency conditions themselves ensure the longevity of the rapid-turn regime. Furthermore, we prove this for the most general definition of rapid turning (i.e., with a turning rate that is large compared to the slow-roll parameters, but not necessarily large compared to unity), using in the process a generalized consistency condition. We also show that a special class of rapid-turn models, including angular inflation, satisfy a large-(compared to $1$)-turn-rate condition even for non-large turning rates.

\end{quotation}

\end{titlepage}

\eject

\tableofcontents

\section{Introduction}
\setcounter{equation}{0}

Models of cosmic acceleration, resulting from the interaction of multiple scalars with gravity, are of great interest both theoretically and phenomenologically. On the theoretical side, they are motivated by recent considerations in the context of compatibility between effective field theories and quantum gravity \cite{OOSV,GK,OPSV,AP,BPR}. And on the phenomenological side, they are significantly richer than the single-scalar models that have been studied extensively in the past. The novel effects arise from background solutions with non-geodesic trajectories in field space, which are characterized by large turning rates. For example, a transient period of rapid turning during inflation can lead to the formation of primordial black holes \cite{PSZ,FRPRW,LA,LA2_pbh}. A sustained rapid-turn phase is also of significant interest \cite{AB,SM,TB,BM,GSRPR,CRS,ACIPWW,ChRoSf,APR2,APR,CR,IMS,ChG}, since it enables slow-roll inflation to occur on steep potentials.\footnote{It is interesting to note that rapid-turn cosmic acceleration can lead to new effects in models of late dark energy too. Notably, such multifield dark energy can have an equation-of-state parameter that is very close to $-1$\,, while being phenomenology distinct from a cosmological constant \cite{ASSV,EASV,ADGW,ADGW2}.}

There are two types of rapid-turn regime, in the inflationary literature, depending on the magnitude of the dimensionless turning rate $\eta_{\perp} \equiv \Omega / H$\,, where $\Omega$ is the turn rate of the field-space trajectory and $H$ - the Hubble parameter of the background space-time solution.\footnote{Here we use the notation $\eta_{\perp}$\,, that is suitable for the two-field case, since we will consider only models with two scalars. In models with more than two scalars, the notation $\omega \equiv \Omega / H$ is often used.} Namely, one type is defined by requiring $\eta_{\perp}^2$ to be large compared to $1$\,, while the other by requiring it to be large compared to the slow-roll parameters. In the second case, $\eta_{\perp}^2$ may even be less than $1$\,. Large values of $\eta_{\perp}^2$ are needed for the primordial-black-hole-generation mechanism mentioned above. However, models of sustained rapid turn inflation, lasting (many) tens of e-folds, can be of either type. In fact, in this context, smaller values of $\eta_{\perp}^2$ may, actually, be preferable since this quantity determines the strength of coupling between certain perturbations, although models with large $\eta_{\perp}^2$ can also be under perturbative control (see \cite{CAP,GSRPR,BFM}). In addition, the inflationary regime with non-large $\eta_{\perp}^2$ may be easier to realize in supergravity, in view of \cite{EGNO,ACPRZ}.

Investigating rapid-turn and third-order slow-roll inflation in two-field models, \cite{AL} found necessary conditions for sustaining this regime at (very) large turning rates. To obtain these conditions, one has to use a basis in field space, which is determined by the scalar potential and field-space metric. This basis is independent of the background trajectory and, at each point of the latter, is related to the more commonly used kinematic basis by a rotation by the so called characteristic angle $\vartheta$\,. These consistency conditions restrict, for a given field-space metric, the form of the potential and/or the relevant part of field space. Also, using generic assumptions, about the components of the Hessian of the potential, seems to suggest that the time-evolution of the characteristic angle is incompatible with a sustained inflationary period of this type, thus calling into question the consistency of rapid turn inflation.

Here we reconsider this problem by using, instead, a rather specific relation between the characteristic angle and the components of the Hessian of the potential, which has to be satisfied during rapid turning. We analyze the more general rapid-turn regime that allows non-large $\eta_{\perp}^2$\,. (Of course, this encompasses the special case of large turning rates.) For that purpose, we use a more general consistency condition, which is easily obtained from the results of \cite{AL}. With the help of the above observation, regarding the special relation between $\vartheta$ and the Hessian of the potential, as well as the generalized consistency condition, we show that the characteristic angle is (nearly-)constant as a function of the number of e-folds, thus ensuring the sustainability of the rapid-turn and slow-roll regime regardless of whether $\eta_{\perp}^2$ is large or non-large. We illustrate this result on the example of side-tracked inflation \cite{GSRPR}. In addition, perhaps surprisingly, we find that a special case of the generalized consistency condition enables certain models with non-large $\eta_{\perp}^2$ to satisfy one of the consistency relations in \cite{AL}, 
which are relevant for very large turning rates. We show that this special case is of crucial importance for the consistency of angular inflation \cite{CRS}.

The organization of this note is the following. In Section \ref{Backgr}, we review the main characteristics of two-field inflationary models. We also introduce the angle of rotation between the kinematic and potential-gradient bases of field space, called characteristic angle. In Section \ref{ConsCond}, we briefly outline the derivation of the consistency conditions of \cite{AL}, relevant for very large turning rates. This enables us to easily obtain a consistency condition for the most general rapid-turn regime, which allows non-large turning rates. We also discuss two important special cases of this general condition, one of which is essential for angular inflation. In Section \ref{SecCharAng}, we point out that, for slow-rolling and rapid-turning inflationary trajectories, there is a specific relation between the characteristic angle and the components of the Hessian of the potential. Using this relation and the generalized consistency condition of the previous section, we show that the characteristic angle remains (nearly-)constant, as a function of the number of e-folds, during rapid-turn and third order slow-roll inflation. Finally, in Section \ref{Examples}, we illustrate our results on two examples of rapid-turn models, namely side-tracked inflation and angular inflation.

\section{Background material} \label{Backgr}
\setcounter{equation}{0}

We will consider multifield inflationary models arising from the interaction of a  number of scalar fields with gravity. Our focus will be on the case with two scalars. The relevant action is: 
\be
\label{Action_gen}
S = \int d^4x \sqrt{-\det g} \left[ \frac{R(g)}{2} - \frac{1}{2} G_{IJ} (\{\varphi^I\}) \pd_{\mu} \varphi^I \pd^{\mu} \varphi^J - V (\{ \varphi^I \}) \right] \,\,\, ,
\ee
where $g_{\mu \nu}$ is the 4d spacetime metric with $\mu,\nu = 0,...,3$\, and $G_{IJ}$ is the metric on the 2d manifold parameterized by the scalars $\{\varphi^I\}$ with $I,J = 1,2$\,. We will assume the standard background Ansatze:
\be
\label{metric_g}
d s^2_g = - d t^2 + a^2(t) \,d \vec{x}^2 \qquad {\rm and} \qquad \varphi^I = \varphi^I (t) \quad .
\ee
Then, one has the following equations of motion:
\be
\label{EinstEqs}
G_{IJ} \dot{\varphi}^I \dot{\varphi}^J = - 2 \dot{H}  \qquad  ,  \qquad  3 H^2 + \dot{H} = V\quad ,
\ee
and
\be \label{EoM_sc}
D_t \dot{\varphi}^I + 3 H \dot{\varphi}^I + G^{IJ} V_J = 0 \quad ,
\ee
where $\dot{} \equiv \pd_t$ and $H = \dot{a} / a$ is the Hubble parameter, $V_J \equiv \pd_{\varphi^J} V$ and we have defined
\be
D_t A^I \equiv \,\dot{\varphi}^J \,\nabla_J A^I = \dot{A}^I + \Gamma^I_{JK}(\varphi) \,\,\dot{\varphi}^J A^K
\ee
for any vector $A^I$ on the manifold parameterized by $\{ \varphi^I \}$ with $\Gamma^I_{JK} (\varphi)$ being the Christoffel symbols of the metric $G_{IJ}$\,.

\subsection{Kinematic basis and inflationary parameters}

A convenient basis in the scalar field-space is given by the tangent and normal vectors to the trajectory $\left( \varphi^1 (t), \varphi^2 (t) \right)$ of a background solution. These vectors can be defined respectively as:
\be \label{N_def}
T^I = \frac{\dot{\varphi}^I}{\dot{\sigma}} \quad {\rm and} \quad N_I = (\det G)^{1/2} \epsilon_{IJ} T^J \quad , \quad {\rm where} \quad \dot{\sigma} \equiv \sqrt{G_{IJ} \dot{\varphi}^I \dot{\varphi}^J} \quad .
\ee
Note that the basis $\left( T(t),N(t) \right)$ is orthonormal. It is useful to project the equations of motion (\ref{EoM_sc}) along the directions of $T^I$ and $N_I$\,, obtaining:
\be
\label{EoM_sigma}
\ddot{\sigma} + 3 H \dot{\sigma} + T^I V_I = 0
\ee
and
\be
\label{EoM_N}
N_I D_t T^I = - \frac{N^I V_I}{\dot{\sigma}} \quad ,
\ee
respectively. Due to (\ref{EoM_N}), the turning rate of a field-space trajectory, which is defined by $\Omega \equiv - N_I D_t T^I$ \cite{AAGP}, can be conveniently rewritten as:
\be
\label{Om_2}
\Omega \,= \frac{N^I V_I}{\dot{\sigma}} \quad .
\ee
This quantity is of utmost importance in rapid turn models, since it measures the deviation of the field-space trajectory from a geodesic. Recall that the qualitatively novel features of the multifield case, compared to the single-field one, arise precisely for strongly non-geodesic trajectories of the background solutions.

The inflationary regime, that we will be interested in, is characterized by the following slow-roll parameters:
\be \label{SR_par}
\varepsilon \equiv -\frac{\dot{H}}{H^2} \qquad , \qquad \eta_{\parallel} \equiv - \frac{\ddot{\sigma}}{H \dot{\sigma}} \qquad , \qquad \xi \equiv \frac{\dddot{\sigma}}{H^2\dot{\sigma}} \quad ,
\ee
as well as the rapid-turn parameters:\footnote{The origin of the notation $\eta_{\parallel,\perp}$ is the following. In the two-field context one can define the analogue of the single-field $\eta$-parameter by $\eta^I \equiv - \frac{1}{H \dot{\sigma}} D_t \dot{\varphi}^I$ \cite{CAP}. Then, decomposing this vector as $\eta^I = \eta_{\parallel} T^I + \eta_{\perp} N^I$ and using (\ref{EoM_sigma})-(\ref{Om_2}), one finds for $\eta_{\parallel}$ and $\eta_{\perp}$ exactly the expressions given in (\ref{SR_par})-(\ref{RT_par}). Alternatively, one can just view (\ref{SR_par})-(\ref{RT_par}) as providing the definitions of all relevant parameters. We have kept the notation $\eta_{\parallel,\perp}$ to facilitate comparison with \cite{AL}.}
\be \label{RT_par}
\eta_{\perp} \equiv \frac{\Omega}{H} \qquad , \qquad \nu \,\equiv \frac{\dot{\eta}_{\perp}}{H \eta_{\perp}} \quad .
\ee
Third order slow-roll inflation is defined by the conditions:
\be \label{SR_3rdO}
\varepsilon <\!\!< 1 \qquad, \qquad |\eta_{\parallel}| <\!\!< 1 \qquad , \qquad |\xi| <\!\!< 1 \quad ,
\ee
which we will assume always in the following. However, there are two types of rapid-turn regime in the literature. One is defined by the stronger condition:
\be \label{RT_s}
\eta_{\perp}^2 >\!\!> 1 \,\,\, ,
\ee
while the other by the weaker requirements:
\be \label{RT_w}
\eta_{\perp}^2 >\!\!> \varepsilon \,, \,|\eta_{\parallel}| \,, \,|\xi| \,\,\, .
\ee
In both cases, one assumes that:
\be \label{nu_small}
|\nu| <\!\!< 1 \,\,\, ,
\ee
in order to sustain the rapid-turning phase for long enough (i.e., for at least 60 or so e-folds of inflation). The studies in \cite{AL} were focused on the stronger condition (\ref{RT_s}). Here we will consider also the more general rapid-turn regime, defined by (\ref{RT_w}). Note that, in this latter case, the magnitude of $\eta_{\perp}$ does not need to be large and, in fact, one may even have $\eta_{\perp}^2 < 1$ in view of (\ref{SR_3rdO}).

Finally, it will be technically useful, in the following, to define the quantity:
\be
\label{c_def}
c \,\equiv \frac{H \dot{\sigma}}{\sqrt{G^{IJ} V_I V_J}} \,\,\, .
\ee
Notice that by definition $c > 0$\,. Also, we will see shortly that the regime (\ref{RT_s}) corresponds to $c^2$ being small. 

\subsection{Potential gradient basis and characteristic angle}

The potential $V (\{ \varphi^I \})$ and metric $G_{IJ} (\{ \varphi^I \})$ determine naturally a basis in field space, which is independent of any inflationary trajectory. Thus, this basis is very useful for studying general properties of the model, without knowing explicit background solutions.\footnote{This potential gradient basis was introduced first in \cite{BM}.} The basis vectors are defined by:
\be
\label{ndef}
n^K = \frac{G^{KL} V_L}{\sqrt{G^{IJ} V_I V_J}} \qquad {\rm and} \qquad \tau_I = (\det G)^{1/2} \epsilon_{IJ} n^J \,\,\, .
\ee
Note that $n$ is a unit vector aligned with the gradient of $V$ and that the basis $\left( n (\varphi) , \tau (\varphi) \right)$ is orthonormal, where we have denoted $\varphi \equiv \{ \varphi^I\}$ for brevity.

At each point of a background trajectory, the basis defined in (\ref{N_def}) is related to the basis in (\ref{ndef}) by the following rotation:
\bea
\label{BasesRel}
T \!&=& \!\cos \vartheta \,\, n \,+ \,\sin \vartheta \,\, \tau \,\,\, , \nn \\
N \!&=& \!-\sin \vartheta \,\, n \,+ \,\cos \vartheta \,\, \tau \,\,\, ,
\eea
where the angle $\vartheta \in (-\pi,\pi]$ depends on time as well as on the trajectory under consideration. Following the terminology of \cite{AL} (although, for convenience, we have simplified the notation), we will call this $\vartheta$ the {\it characteristic angle} of a field-space trajectory. Using the definitions in (\ref{SR_par})-(\ref{RT_par}) and equations (\ref{EoM_sigma}) and (\ref{Om_2}), one can show that \cite{AL}:
\be \label{eta_cos_sin_c}
\eta_{\parallel} = 3 + \frac{\cos \vartheta}{c} \qquad , \qquad \eta_{\perp} = - \frac{\sin \vartheta}{c} \quad ,
\ee
where we have also used (\ref{BasesRel}) and (\ref{c_def}). Note that these relations imply:
\be \label{c_exact}
\eta_{\perp}^2 + (\eta_{\parallel} - 3)^2 = \frac{1}{c^2} \qquad {\rm and} \qquad \eta_{\perp}^2 \,= \,(\eta_{\parallel} - 3)^2 \tan^2 \vartheta \quad .
\ee

For future use, let us also point out that, during slow-roll, (\ref{eta_cos_sin_c}) leads to:
\be \label{cos_sin_SR}
\cos \vartheta \cong - 3 c \qquad {\rm and\,\,thus} \qquad \sin \vartheta \cong s \sqrt{1-9c^2} \quad , \quad {\rm where} \quad s = \pm 1 \quad .
\ee
Similarly, for $|\eta_{\parallel}| <\!\!< 1$\,, we find from the two relations in (\ref{c_exact}):
\be \label{eta_perp_SR}
\eta_{\perp}^2 \cong \frac{1}{c^2} - 9 \qquad {\rm and} \qquad \eta_{\perp}^2 \cong 9 \,\tan^2 \vartheta \quad ,
\ee
respectively. Notice that we have introduced special notation, namely ``$\cong$", for approximate equalities that rely only on the slow-roll approximation. Since, in the following, we will be interested in distinguishing between the two types of rapid turning given by (\ref{RT_s}) and (\ref{RT_w}), we will use from now on the notation ``$\simeq$" for approximate equalities satisfied in the regime (\ref{RT_w}), but which do not need (\ref{RT_s}). And, finally, we will reserve the symbol ``$\approx$" for approximate equalities that require (\ref{RT_s}).

From (\ref{eta_perp_SR}), we can see that, during slow-roll, large $\eta_{\perp}^2$ corresponds to small $c^2$ and also that large variations of $\vartheta$ imply large variations of $\eta_{\perp}^2$\,, in potential conflict with the requirement (\ref{nu_small}). Based on genericity arguments, it would seem that the evolution of $\vartheta (t)$ leads to a fast (i.e. within a rather limited number of e-folds) exit from the rapid-turning phase \cite{AL}. Our goal here will be to show that, in fact, this does not occur due to important specifics of the rapid-turn regime. More precisely, we will prove that the relevant consistency conditions guarantee a (near-)constant $\vartheta$, thus ensuring the longevity of the rapid-turning phase.

\section{Rapid-turn consistency conditions} \label{ConsCond}
\setcounter{equation}{0}

We begin by briefly reviewing the main steps in the derivation of the rapid-turn consistency conditions of \cite{AL}, relevant for the regime (\ref{RT_s}). This will be useful not only for more clarity of the present exposition, but also because it will enable us to easily write down the more general consistency condition that is valid in the regime (\ref{RT_w}).

The starting point is the exact relations \cite{HP,AGHPP,CCLBNZ}:
\bea
\frac{V_{TT}}{3 H^2} &=& \frac{\eta_{\perp}^2}{3} + \varepsilon + \eta_{\parallel} - \frac{\xi}{3} \,\,\, , \nn \\
\frac{V_{TN}}{H^2} &=& \eta_{\perp} \left( 3 - \varepsilon - 2 \eta_{\parallel} + \nu \right) \,\,\, ,
\eea
which follow from (\ref{EoM_sigma}) and (\ref{Om_2}); here $V_{TT} \equiv T^I T^J \nabla_I V_J$ and $V_{TN} \equiv T^I N^J \nabla_I V_J$. We will always assume that (\ref{SR_3rdO}) and (\ref{nu_small}) are satisfied. Hence, in either of the approximations (\ref{RT_s}) or (\ref{RT_w}), we have:
\be \label{TwoRel}
\frac{V_{TT}}{3 H^2} \simeq \frac{\eta_{\perp}^2}{3} \qquad {\rm and} \qquad \frac{V_{TN}}{H^2} \simeq 3 \eta_{\perp} \,\,\, .
\ee
The use of the symbol ``$\simeq$" here is meant, as explained above, to underline that these approximate relations rely on (\ref{RT_w}), but do not require (\ref{RT_s}), and thus are valid even for non-large $\eta_{\perp}^2$\,. Now, introducing the notation $V_{nn} \equiv n^I n^J \nabla_I V_J$\,, $V_{n \tau} = n^I \tau^J \nabla_I V_J$\,, $V_{\tau \tau} \equiv \tau^I \tau^J \nabla_I V_J$ and using (\ref{BasesRel}), one can write:
\bea
\label{VTN-Vntau}
\hspace*{-0.4cm}V_{TT} &=& V_{nn} \,\cos^2 \vartheta \,+ \,2 \,V_{n \tau} \,\sin \vartheta \,\cos \vartheta \,+ \,V_{\tau \tau} \,\sin^2 \vartheta \,\,\, , \nn \\
\hspace*{-0.4cm}V_{TN} &=& \left( V_{\tau \tau} - V_{nn} \right) \,\cos \vartheta \,\sin \vartheta \,+ \,V_{n \tau} \left( \cos^2 \vartheta - \sin^2 \vartheta \right) \,\, .
\eea
Substituting (\ref{VTN-Vntau}) in (\ref{TwoRel}) and using (\ref{eta_cos_sin_c}), together with (\ref{cos_sin_SR}) and the slow-roll relation $3H^2 \cong V$, gives:
\bea
\label{VTN-Vntau_V_c}
\frac{V}{3\,c^2} - 3 V &\simeq & 9 c^2 V_{nn} - 6 s c \,\sqrt{1-9c^2} \,V_{n \tau} + (1-9c^2) V_{\tau \tau} \,\,\, , \nn \\
\frac{s}{c} \sqrt{1-9c^2}\,V &\simeq & 3 s c \,\sqrt{1-9c^2} \left( V_{\tau \tau} - V_{nn} \right) + (1-18c^2) V_{n \tau} \,\,\, .
\eea
Now, the basic idea is to eliminate $c$ from this system, in order to be left with a relation only between $V$ and its Hessian components. For that purpose, after some manipulations of (\ref{VTN-Vntau_V_c}), one can find the following relation \cite{AL}:\footnote{More precisely, one can solve the second equation in (\ref{VTN-Vntau_V_c}) for $V_{n \tau}$ and substitute the result in the first equation of (\ref{VTN-Vntau_V_c}). Then, the latter can be solved for $V_{\tau \tau}$ in terms of $c$, $V$ and $V_{nn}$\,. Substituting this $V_{\tau \tau}$-expression in the expression for $V_{n \tau}$\,, obtained in the first step described here, one finds (\ref{Vnt_c_Vnn}); in particular, the $V$-terms cancel exactly.}
\be \label{Vnt_c_Vnn}
V_{n \tau} \,\simeq \frac{3sc V_{nn}}{\sqrt{1-9c^2}} \,\,\, .
\ee
It merits underlining that the steps, which lead from (\ref{VTN-Vntau_V_c}) to (\ref{Vnt_c_Vnn}), do not use {\it any} approximation; the approximate equality sign ``$\simeq$" in (\ref{Vnt_c_Vnn}) is only due to the same signs in (\ref{VTN-Vntau_V_c}). 

In the rapid-turn approximation (\ref{RT_s}), that was the focus of \cite{AL}, we have $c^2 <\!\!< 1$ due to (\ref{eta_perp_SR}). Hence, (\ref{VTN-Vntau_V_c}) simplifies to:
\bea \label{Apr_Eqs}
\frac{V}{3\,c^2} &\approx & 9 c^2 V_{nn} - 6 s c \,V_{n \tau} + V_{\tau \tau} \,\,\, , \nn \\
\frac{s V}{c} &\approx & 3 s c \,\left( V_{\tau \tau} - V_{nn} \right) + V_{n \tau} \,\,\, ,
\eea
whereas from (\ref{Vnt_c_Vnn}) we obtain:
\be \label{c_apr}
c \,\approx \,\frac{s}{3} \frac{V_{n \tau}}{V_{nn}} \,\,\, .
\ee
Using (\ref{c_apr}) in the system (\ref{Apr_Eqs}), the second equation gives: 
\be \label{E2_apr}
3V V_{nn}^2 \,\approx \,V_{n \tau}^2 V_{\tau \tau} \,\,\, ,
\ee
while the first equation leads to:
\be \label{E1_apr}
\frac{3V V_{nn}^2}{V_{n \tau}^2} \,\approx \,V_{\tau \tau} - \frac{V_{n \tau}^2}{V_{nn}} \,\,\, .
\ee
Comparing (\ref{E2_apr}) and (\ref{E1_apr}), we conclude that to have an approximate solution of the system (\ref{Apr_Eqs}), the following inequality has to be satisfied:
\be \label{Cons_Ineq}
V_{\tau \tau} |V_{nn}| \,>\!\!> \,V_{n \tau}^2 \,\,\, ,
\ee
where we have taken into account that $V_{\tau \tau} > 0$ due to (\ref{E2_apr}). The pair of consistency conditions (\ref{E2_apr}) and (\ref{Cons_Ineq}) was obtained in \cite{AL}; see also the concise summary in \cite{AL2}.

Note, however, that (\ref{Vnt_c_Vnn}) is easily solved for $c$ without assuming $c^2 <\!\!< 1$\,. Namely, we obtain:
\be \label{c_sol}
c^2 \simeq \frac{V_{n \tau}^2}{9 (V_{n \tau}^2 + V_{nn}^2)} \,\,\, .
\ee
In view of (\ref{c_def}), which implies that $c > 0$ always, we have from (\ref{c_sol}):
\be \label{c_abs}
c \simeq \frac{|V_{n \tau}|}{3 \sqrt{V_{n \tau}^2 + V_{nn}^2}} \,\,\, .
\ee
Since this result relies only on the approximations (\ref{RT_w}), but not on (\ref{RT_s}), it enables us to obtain consistency conditions valid for non-large $\eta_{\perp}^2$\,. In fact, both equations in the system (\ref{VTN-Vntau_V_c}) lead to the same condition. Indeed, substituting (\ref{c_abs}), we find from the first and second equations of (\ref{VTN-Vntau_V_c}) respectively:
\be \label{Eq1_full_c}
\frac{3V V_{nn}^2}{V_{n \tau}^2} \,\simeq \,\frac{V_{n \tau}^2 V_{nn}}{(V_{n \tau}^2 + V_{nn}^2)} - \frac{2 s |V_{n \tau}| |V_{nn}| V_{n \tau}}{(V_{n \tau}^2+V_{nn}^2)} + \frac{V_{nn}^2 V_{\tau \tau}}{(V_{n \tau}^2+V_{nn}^2)}
\ee
and
\be \label{Eq2_full_c}
\frac{3s |V_{nn}| V}{|V_{n \tau}|} \,\simeq \,\frac{s |V_{n \tau}| |V_{nn}|}{(V_{n \tau}^2 + V_{nn}^2)} (V_{\tau \tau} - V_{nn}) + \frac{(V_{nn}^2 - V_{n \tau}^2)}{(V_{nn}^2 + V_{n \tau}^2)} V_{n \tau} \,\,\, .
\ee
Now, notice that (\ref{Vnt_c_Vnn}) implies:
\be \label{sgns}
{\rm sign}(V_{n \tau}) = {\rm sign}(sV_{nn}) \,\,\, ,
\ee
due to $c > 0$\,. Hence $s |V_{nn}|/|V_{n \tau}| = V_{nn}/V_{n \tau}$ and $s|V_{n \tau}| |V_{nn}| = V_{n \tau} V_{nn}$\,. Thus, one can see that both (\ref{Eq1_full_c}) and (\ref{Eq2_full_c}) give the same condition:
\be \label{CC_full}
3V V_{nn} (V_{nn}^2 + V_{n \tau}^2) \simeq V_{n \tau}^2 (V_{nn} V_{\tau \tau} - V_{n \tau}^2) \,\,\, .
\ee
To recapitulate, (\ref{CC_full}) is the single consistency condition for the more general rapid-turn regime (\ref{RT_w}), in which one can have non-large $\eta_{\perp}^2$\,. 

It should be noted that this condition is contained implicitly in \cite{TB}. Indeed, equating the two expressions for the dimensionless turning rate, given in eqs. (13) there (as well as identifying $v$ and $w$ in \cite{TB} with our $n$ and $\tau$, respectively), leads precisely to our relation (\ref{CC_full}). However, from those considerations it is not immediately clear whether there are no additional, subleading for large turn rate, contributions to the expressions in eqs. (13) of \cite{TB}.\footnote{Such contributions have been neglected in preceding steps. For instance, one can compute that the more precise form of equation (11) in \cite{TB} is the following: $\frac{\omega}{\omega^2+9} \left( \frac{V_{vw}}{H^2}-3\frac{V_{vv}}{\omega H^2} \right) = {\cal O} (\varepsilon)$.} Also, there is no mention of the third-order slow-roll parameter $\xi$ there. On the other hand, our derivation of the consistency condition (\ref{CC_full}) makes clear that it is valid for any value of $\eta_{\perp}$ satisfying (\ref{RT_w}), as well as for third-order slow roll. Furthermore, (some of) the intermediate results in our derivation will be needed in the subsequent considerations here.\footnote{The condition (\ref{CC_full}) was obtained also in \cite{WIA}, which came out simultaneously with the present paper. It merits underlining that this consistency condition is only necessary, but may not be sufficient, for the existence of solutions to the equations of motion in the rapid-turn and slow-roll regime. Indeed, although (\ref{CC_full}) follows from the field equations and the relevant approximations, the converse is not true. So, even though this relation is useful for selecting suitable models and appropriate parts of their field (and/or parameter) spaces, one should still check whether the equations of motion and needed approximations are satisfied. This task is facilitated by the additional condition on $V$ and its Hessian components, derived in \cite{WIA}, that ensures $\varepsilon <\!\!< 1$. It would be interesting to explore whether the rest of the approximations in (\ref{SR_3rdO}) and (\ref{RT_w})-(\ref{nu_small}) can be encoded by similar conditions on $V$ and $G_{IJ}$\,. However, it seems unlikely that the equations of motion can be entirely substituted, in general, solely by conditions on $V$ and $G_{IJ}$\,, although it would certainly be very interesting if that were proven to be the case.}

Note that using (\ref{c_sol}) in (\ref{eta_perp_SR}) gives:
\be \label{eta_perp_simeq}
\eta_{\perp}^2 \cong \frac{1}{c^2} - 9 \,\simeq \,9 \,\frac{V_{nn}^2}{V_{n \tau}^2} \,\,\, .
\ee
So requiring that:\footnote{Actually, the approximation (\ref{RT_Vnn_Vnt_cond}) can be called `very rapid' turning, since having $\frac{V_{nn}^2}{V_{n \tau}^2} \sim {\cal O} (10)$ is enough to ensure that (\ref{eta_perp_simeq}) is large.}
\be \label{RT_Vnn_Vnt_cond}
V_{nn}^2 >\!\!> V_{n \tau}^2
\ee
ensures that (\ref{RT_s}) is satisfied.

When (\ref{RT_Vnn_Vnt_cond}) and (\ref{Cons_Ineq}) are satisfied, it is easy to see that (\ref{CC_full}) reduces to (\ref{E2_apr}), as well as that (\ref{c_sol}) simplifies to (\ref{c_apr}). Also, using (\ref{E2_apr}) in (\ref{eta_perp_simeq}), we obtain:
\be
\eta_{\perp}^2 \,\approx \,3 \,\frac{V_{\tau \tau}}{V} \,\,\, ,
\ee
which implies that, in this regime, one has $V_{\tau \tau} >\!\!> V$\,. Notice, however, that the latter condition is not necessary in the rapid-turn regime (\ref{RT_w}) with non-large $\eta_{\perp}^2$\,.

For future use, let us note that using (\ref{CC_full}) in (\ref{c_sol}) gives:
\be \label{csq_full_rel}
c^2 \,\simeq \,\frac{V V_{nn}}{3 \,(V_{nn} V_{\tau \tau} - V_{n \tau}^2)} \,\,\, .
\ee
In the approximation (\ref{Cons_Ineq}), this reduces to:
\be \label{c_large_eta}
c^2 \,\approx \,\frac{V}{3 \,V_{\tau \tau}} \,\,\, .
\ee
From (\ref{csq_full_rel}) we also see that, when $V_{nn} > 0$\,, we must have $V_{nn} V_{\tau \tau} > V_{n \tau}^2$\,, even though in the regime (\ref{RT_w}) the strong inequality (\ref{Cons_Ineq}) is not needed.

\subsection{Special case with $V_{nn} \simeq - V_{\tau \tau}$} \label{Vnn_eq_m_Vtt}

There is an interesting special case, in which a rapid-turn model with non-large $\eta_{\perp}^2$ satisfies, in fact, the (very) large-$\eta_{\perp}^2$ consistency condition (\ref{E2_apr}), while (\ref{Cons_Ineq}) may be violated. This case turns out to be of crucial importance for the consistency of angular inflation, as we will discuss in Subsection \ref{AngInfl}. It is characterized by:
\be \label{Vnn_m_Vtt}
V_{nn} \simeq - V_{\tau \tau} \,\,\, .
\ee
In view of (\ref{E2_apr}) implying $V_{\tau \tau} > 0$\,, the relation (\ref{Vnn_m_Vtt}) means that $V_{nn} < 0$ in such a model. Recall, however, that this does not signify instability in the rapid turn context \cite{ChRS}.

To understand this special case, let us substitute (\ref{Vnn_m_Vtt}) inside (\ref{CC_full}) and rewrite the result in the following way:
\be \label{Quad_Eq_Vnt}
V_{n \tau}^4 + (V_{\tau \tau}^2 - 3V V_{\tau \tau}) V_{n \tau}^2 - 3 V V_{\tau \tau}^3 \simeq 0 \,\,\, .
\ee
Now, we can view this as a quadratic equation for $V_{n \tau}^2$\,. One of its two roots, namely $V_{n \tau}^2 \simeq - V_{\tau \tau}^2$\,, is obviously not physical, since it would imply that $V_{n \tau}$ is not real. However, the other root of (\ref{Quad_Eq_Vnt}), given by:
\be \label{CC_Vnn_m_Vtt}
V_{n \tau}^2 \simeq 3 V V_{\tau \tau} \,\,\, ,
\ee
is physically allowed. Furthermore, note that substituting (\ref{Vnn_m_Vtt}) inside (\ref{E2_apr}) gives precisely (\ref{CC_Vnn_m_Vtt}). So a rapid-turn model, with the property (\ref{Vnn_m_Vtt}) along the inflationary trajectory, satisfies (\ref{E2_apr}) regardless of whether or not $\eta_{\perp}^2$ is large compared to $1$\,. And this is true even when the additional large-$\eta_{\perp}^2$ condition (\ref{Cons_Ineq}) is not satisfied.

We will see in Subsection \ref{AngInfl} that this conclusion is essential for the consistency of angular inflation \cite{CRS}. This is because in that model (\ref{E2_apr}) is satisfied, while (\ref{Cons_Ineq}) is problematic. However, along the inflationary trajectory in angular inflation one has precisely the relation (\ref{Vnn_m_Vtt}), thus ensuring the consistency of this rapid-turn solution. 

Finally, note that substituting (\ref{Vnn_m_Vtt}) in (\ref{csq_full_rel}) gives:
\be \label{c_Vnn_min_Vtt}
c^2 \simeq \frac{V V_{\tau \tau}}{3 (V_{\tau \tau}^2 + V_{n \tau}^2)} \,\,\, .
\ee
This reduces to (\ref{c_large_eta}) when $V_{\tau \tau}^2 >\!\!> V_{n \tau}^2$\,, as should be the case. However, for models satisfying (\ref{Vnn_m_Vtt}) along the inflationary trajectory, the condition $V_{\tau \tau}^2 >\!\!> V_{n \tau}^2$ is not necessary for the consistency of the rapid-turn regime, as explained above. Hence, for such models, (\ref{c_Vnn_min_Vtt}) may give results that are significantly different from (\ref{c_large_eta}).

\subsection{Special case with $V_{n \tau}\,, V_{nn} \simeq 0$} \label{VntVnn_0}

Another special case (already pointed out in \cite{AL}), which merits discussing, occurs when:
\be \label{VnnVnt_0}
V_{nn}\,, V_{n \tau} \,\simeq \,0 \,\,\,\, .
\ee
This is relevant, for example, for hyperinflation \cite{AB} since in that case $V_{n \tau} = 0$ identically; see, for instance, the Appendix of \cite{AL2}. Notice from (\ref{Vnt_c_Vnn}) that $V_{n \tau} \simeq 0$ implies $V_{nn} \simeq 0$ and vice versa. One can also verify this directly from (\ref{VTN-Vntau_V_c}). Clearly, inflationary models with the property (\ref{VnnVnt_0}) satisfy trivially both the generalized condition (\ref{CC_full}), valid even for non-large $\eta_{\perp}^2$\,, as well as the (very) large-$\eta_{\perp}^2$ condition (\ref{E2_apr}), although not (\ref{Cons_Ineq}). 

Substituting (\ref{VnnVnt_0}) in the second equation of (\ref{VTN-Vntau_V_c}) gives:
\be \label{c_Vnn_zero}
c^2 \,\simeq \,\frac{V}{3 \,V_{\tau \tau}} \,\,\,\, .
\ee
It is easy to see that (\ref{c_Vnn_zero}) solves also the first equation of (\ref{VTN-Vntau_V_c}), without the use of any approximation. Note that this is the same expression as (\ref{c_large_eta}), although now we have not relied on (\ref{RT_s}). This suggests that, in the special case with $V_{nn}\,, V_{n \tau} \simeq 0$\,\,, the rapid-turn regime with non-large-$\eta_{\perp}^2$ is not qualitatively different from the (very) large-$\eta_{\perp}^2$ regime. Hence, inflationary solutions, satisfying (\ref{VnnVnt_0}) along their field-space trajectories, could provide valuable toy models for studying the rapid-turn regime with non-large-$\eta_{\perp}^2$\,, in technically much simpler settings than in generic models of the type (\ref{RT_w}).

\section{Evolution of the characteristic angle} \label{SecCharAng}
\setcounter{equation}{0}

Now we are ready to investigate the evolution of the characteristic angle $\vartheta$ of the inflationary field-space trajectories during a rapid-turning phase. Recall that this angle, introduced in (\ref{BasesRel}), relates the kinematic and the potential gradient bases in field space to each other. One can obtain an equation, determining the function $\vartheta (t)$, by projecting (\ref{EoM_sc}) along either of the vectors $n$ or $\tau$ \cite{AL}; both projections lead to the same equation, upon using (\ref{c_def}) and the first relation in (\ref{eta_cos_sin_c}). In the present Section, we will show how the solution of this equation is compatible with a sustained rapid-turn period. More precisely, we will see that, due to the consistency conditions of Section \ref{ConsCond}, the characteristic angle $\vartheta$ is (nearly-)constant, as a function of the number of e-folds, in either of the rapid-turn regimes (\ref{RT_s}) and (\ref{RT_w}).

\subsection{General case with $V_{nn}\,, V_{n \tau} \neq 0$}

In the slow roll approximation, the angle $\vartheta$ is determined by the following equation (see Appendix \ref{Deriv_Eq_Th}):
\be \label{EoMTh_ph_2}
\frac{d\vartheta}{dN} + 3 \tan \vartheta - \frac{1}{V} \left( V_{n \tau} \cos^2 \vartheta + V_{\tau \tau} \cos \vartheta \sin \vartheta \right) \,\cong \,0 \,\,\,\, ,
\ee
where \,$N \equiv \int H dt$ \,is the number of e-folds of inflationary expansion. To analyze the implications of (\ref{EoMTh_ph_2}), note that combining (\ref{cos_sin_SR}) with (\ref{c_abs}) gives: 
\be \label{cos_f}
\cos \vartheta \,\cong - 3 c \,\simeq - \frac{|V_{n \tau}|}{\sqrt{V_{n \tau}^2 + V_{nn}^2}}
\ee
and
\be \label{sin_f}
\sin \vartheta \,\cong \,s \sqrt{1 - 9c^2} \,\simeq \frac{s |V_{nn}|}{\sqrt{V_{n \tau}^2+V_{nn}^2}} \,\,\,\, .
\ee 
Substituting (\ref{cos_f})-(\ref{sin_f}) in (\ref{EoMTh_ph_2}), we obtain:
\be \label{Th_eq_nl}
\frac{d\vartheta}{dN} - 3 \frac{s |V_{nn}|}{|V_{n \tau}|} - \frac{(V_{n \tau}^3 - V_{\tau \tau} |V_{n \tau}| s |V_{nn}|)}{V (V_{n \tau}^2 + V_{nn}^2)} \,\simeq \,0 \,\,\,\, .
\ee
Due to (\ref{sgns}), we can simplify (\ref{Th_eq_nl}) as:
\be \label{Th_eq_full}
\frac{d \vartheta}{d N} \,\simeq \,\frac{ 3V V_{nn} (V_{n \tau}^2 + V_{nn}^2) + V_{n \tau}^4 - V_{\tau \tau} V_{nn} V_{n \tau}^2 }{V V_{n \tau} (V_{n \tau}^2+V_{nn}^2)} \,\,\,\, .
\ee 
Now note that the expression in the numerator of (\ref{Th_eq_full}) is precisely the combination that vanishes due to (\ref{CC_full}). Hence we have:
\be \label{Th_const}
\frac{d \vartheta}{d N} \simeq 0 \,\,\,\, .
\ee
We should underline that this is valid in the most general rapid-turn regime (\ref{RT_w}), regardless of whether $\eta_{\perp}^2$ is large or not. In view of (\ref{eta_perp_SR}), the result (\ref{Th_const}) ensures that (\ref{nu_small}) is satisfied and thus the rapid-turning phase is sustainable for a large number of e-folds.\footnote{Note that we cannot conclude from (\ref{Th_const}) that $\nu$ vanishes, since there may be subleading corrections proportional to (powers of) the slow roll parameters which we have been consistently neglecting. In fact, even without taking into account such corrections to (\ref{Th_const}), one can see that during slow roll $\nu$ is small, but non-vanishing. Namely, substituting the second relation of (\ref{c_exact}) inside (\ref{RT_par}), we find that $\nu = \frac{1}{(\eta_{\parallel} - 3)} \frac{d \eta_{\parallel}}{dN} + \frac{2}{\sin (2 \vartheta)} \frac{d \vartheta}{dN}$\,. Now, from (\ref{SR_par}) one can compute: $\frac{\dot{\eta}_{\parallel}}{H} = - \frac{\dddot{\sigma}}{\dot{\sigma} H^2} + \left( \frac{\ddot{\sigma}}{\dot{\sigma} H} \right)^{\!2} + \frac{\ddot{\sigma} \dot{H}}{\dot{\sigma} H^3} = - \xi + \eta_{\parallel}^2 + \varepsilon \eta_{\parallel}$\,. Thus, even for vanishing $\frac{d\vartheta}{dN}$\,, one has $\nu \cong \frac{1}{3} (\xi - \eta_{\parallel}^2 - \varepsilon \eta_{\parallel})$\,, implying that $|\nu|$ is small but nonzero.}

It is instructive to see how the above computation specializes in the very rapid-turn regime. In that case, the first term in the combination:
\be
V_{n \tau} \cos^2 \vartheta \,+ \,V_{\tau \tau} \cos \vartheta \sin \vartheta \,\,\, ,
\ee
that appears in (\ref{EoMTh_ph_2}), is negligible compared to the second. Indeed, let us consider the magnitude of the ratio of these two terms:
\be \label{Comp2Ts}
\bigg| \frac{V_{n \tau}}{V_{\tau \tau}} \cot \vartheta \,\bigg| \,\approx \,\frac{V_{n \tau}^2}{V_{\tau \tau} |V_{nn}|} \,<\!\!< \,1 \,\,\, ,
\ee
where we have used that (\ref{cos_f}) and (\ref{sin_f}) simplify to 
\be \label{cos_sin_s}
\cos \vartheta \approx - \frac{|V_{n \tau}|}{|V_{nn}|} \qquad {\rm and} \qquad \sin \vartheta \,\approx \,s \,\,\,\, ,
\ee 
respectively, due to (\ref{RT_Vnn_Vnt_cond}); the last step in (\ref{Comp2Ts}), clearly, relies on (\ref{Cons_Ineq}). So in this case the $\vartheta$-equation (\ref{EoMTh_ph_2}) reduces to: 
\be \label{EoMTh_s}
\frac{d \vartheta}{d N} + \tan \vartheta - \frac{V_{\tau \tau} \cos \vartheta \sin \vartheta}{3V} \,\approx \,0 \,\,\,\, .
\ee
Of course, substituting (\ref{cos_sin_s}) in (\ref{EoMTh_s}) and using (\ref{E2_apr}), we find again that $d \vartheta / dN \approx 0$\,, in accord with (\ref{Th_const}). However, the simplified $\vartheta$-equation (\ref{EoMTh_s}) suggests, again, a qualitative similarity between the (very) large-$\eta_{\perp}^2$ regime and the special case of (\ref{RT_w}), which was considered in Subsection \ref{VntVnn_0}. In fact, the latter case merits a separate (albeit brief) discussion, since $c$ is given by a different expression than (\ref{c_abs}), that was used above.

\subsection{Special case with $V_{n \tau}\,, V_{nn} \simeq 0$}

Recall that, in this case, the expression for $c$ is given by (\ref{c_Vnn_zero}). Using that in (\ref{cos_sin_SR}), we have:
\be \label{ThEq_Vnt_0}
\cos \vartheta \,\simeq - \sqrt{\frac{3 V}{V_{\tau \tau}}} \quad .
\ee
Since $V_{n \tau} \simeq 0$\,, the $\vartheta$-equation (\ref{EoMTh_ph_2}) reduces again to the form in (\ref{EoMTh_s}), which can be rewritten conveniently as:
\be \label{ThEq_Vnt_0c}
\frac{d \vartheta}{d N} + \tan \vartheta \left( 1 - \frac{V_{\tau \tau} \cos^2 \vartheta }{3V} \right) \,\simeq \,0 \,\,\,\, .
\ee
Obviously, substituting (\ref{ThEq_Vnt_0}) in (\ref{ThEq_Vnt_0c}) leads to (\ref{Th_const}), yet again.

\section{Examples} \label{Examples}
\setcounter{equation}{0}

In this section we illustrate our considerations on two examples of rapid turn inflation. The first example, side-tracked inflation \cite{GSRPR}, allows for both regimes (\ref{RT_s}) and (\ref{RT_w}), and thus is very suitable for illustrating (\ref{Th_const}). The second example, angular inflation \cite{CRS}, underscores the importance of the special case we considered in Subsection \ref{Vnn_eq_m_Vtt}.

\subsection{Side-tracked inflation} \label{SubSec-Side-tr}

Side-tracked inflation was introduced in \cite{GSRPR}. In its minimal realization, the field-space metric $G_{IJ}$ and scalar potential $V$ have the form:
\be \label{Side-tr_GV}
ds^2_G = \left( 1 + \frac{2 \chi^2}{M^2} \right) d\phi^2 + d \chi^2 \qquad , \qquad V (\phi, \chi) = U (\phi) + \frac{m_h^2}{2} \chi^2 \,\,\,\, .
\ee
Important characteristics of the model are that the field $\chi$ is heavy, i.e. its mass satisfies $m_h >\!\!> H$\,, and that the potential $V$ is dominated by $U$ along the inflationary field-space trajectory. The latter is given by \cite{GSRPR}:
\be \label{chi_traj}
\chi^2 \simeq \frac{M^2}{2} \left( \sqrt{\frac{2}{3}} \,\frac{|U'|}{m_h M \sqrt{U}} \,- \,1 \right) \,\,\, ,
\ee
where we have denoted $' \equiv \frac{d}{d\phi}$\,. One also has \cite{GSRPR}:
\be \label{dPhdN}
\frac{d \phi}{dN} \simeq - \sqrt{\frac{3}{2}} \,\frac{m_h M}{\sqrt{V}} \,{\rm sign} (U') \,\,\, ,
\ee
with $N$ being the number of e-folds.

To compute the components of the Hessian of $V$, it is convenient to introduce the notation:
\be \label{nt_hat_def}
n^I \equiv \frac{\hat{n}^I}{\hat{N}} \qquad {\rm and} \qquad \tau^I \equiv \frac{\hat{\tau}^I}{\hat{N}} \quad ,
\ee
where $\hat{N}$ was defined in (\ref{Nhat_V}). Then, in terms of \,$\hat{V}_{nn} \equiv \hat{n}^I \hat{n}^J \nabla_I V_J$\,\,, $\hat{V}_{n \tau} \equiv \hat{n}^I \hat{\tau}^J \nabla_I V_J$ and \,$\hat{V}_{\tau \tau} \equiv \hat{\tau}^I \hat{\tau}^J \nabla_I V_J$\,\,, one has the relations:
\be \label{Vh_Nh_V}
\hat{V}_{nn} = V_{nn} \hat{N}^2 \quad , \quad \hat{V}_{n \tau} = V_{n \tau} \hat{N}^2 \quad , \quad \hat{V}_{\tau \tau} = V_{\tau \tau} \hat{N}^2 \quad .
\ee
Using (\ref{ndef}) and (\ref{Side-tr_GV}), one finds (see the Appendix of \cite{AL2}):
\bea
\hat{V}_{nn} &=& \frac{U'^2 \!\left( U'' - \frac{2 m_h^2 \chi^2}{M^2} \right)}{\left( 1 + \frac{2 \chi^2}{M^2} \right)^{\!2}} + m_h^6 \chi^2 \quad , \label{V_h_nn} \nn \\
\hat{V}_{n \tau} &=& 
\frac{2 \,\chi \,U'^3}{M^2 \!\left( 1 + \frac{2 \chi^2}{M^2} \right)^{\!5/2}} + \frac{m_h^2 \,\chi \,U' \,U''}{\left( 1 + \frac{2 \chi^2}{M^2} \right)^{\!3/2}} - \frac{m_h^4 \,\chi \,U'}{\left( 1 + \frac{2 \chi^2}{M^2} \right)^{\!1/2}} \quad , \nn \\
\hat{V}_{\tau \tau} &=& \frac{m_h^2 \,U'^2 \!\left( 1 + \frac{6 \chi^2}{M^2} \right)}{\left( 1 + \frac{2 \chi^2}{M^2} \right)^2} + \frac{m_h^4 \chi^2 \left( U'' + \frac{2 m_h^2 \chi^2}{M^2} \right)}{\left( 1 + \frac{2 \chi^2}{M^2} \right)} \label{V_h_tt} \quad ,
\eea
as well as:
\be \label{Side-tr_N}
\hat{N}^2 = \frac{U'^2}{1+\frac{2 \chi^2}{M^2}} + m_h^4 \chi^2 \quad .
\ee
Substituting (\ref{chi_traj}) in (\ref{V_h_tt})-(\ref{Side-tr_N}) and using (\ref{Vh_Nh_V}), one can obtain $V_{nn}$\,, $V_{n\tau}$ and $V_{\tau \tau}$ in terms of $U$ and its derivatives. Using those (cumbersome) expressions, it was shown analytically in the Appendix of \cite{AL2} that side-tracked inflation satisfies (\ref{E2_apr}) in the limit of very rapid turning. The general case, defined by (\ref{RT_w}), seems too messy to analyze analytically, although of course one can verify (\ref{CC_full}) on numerical examples. 

\begin{figure}[t]
\begin{center}
\hspace*{-0.6cm}
\includegraphics[scale=0.344]{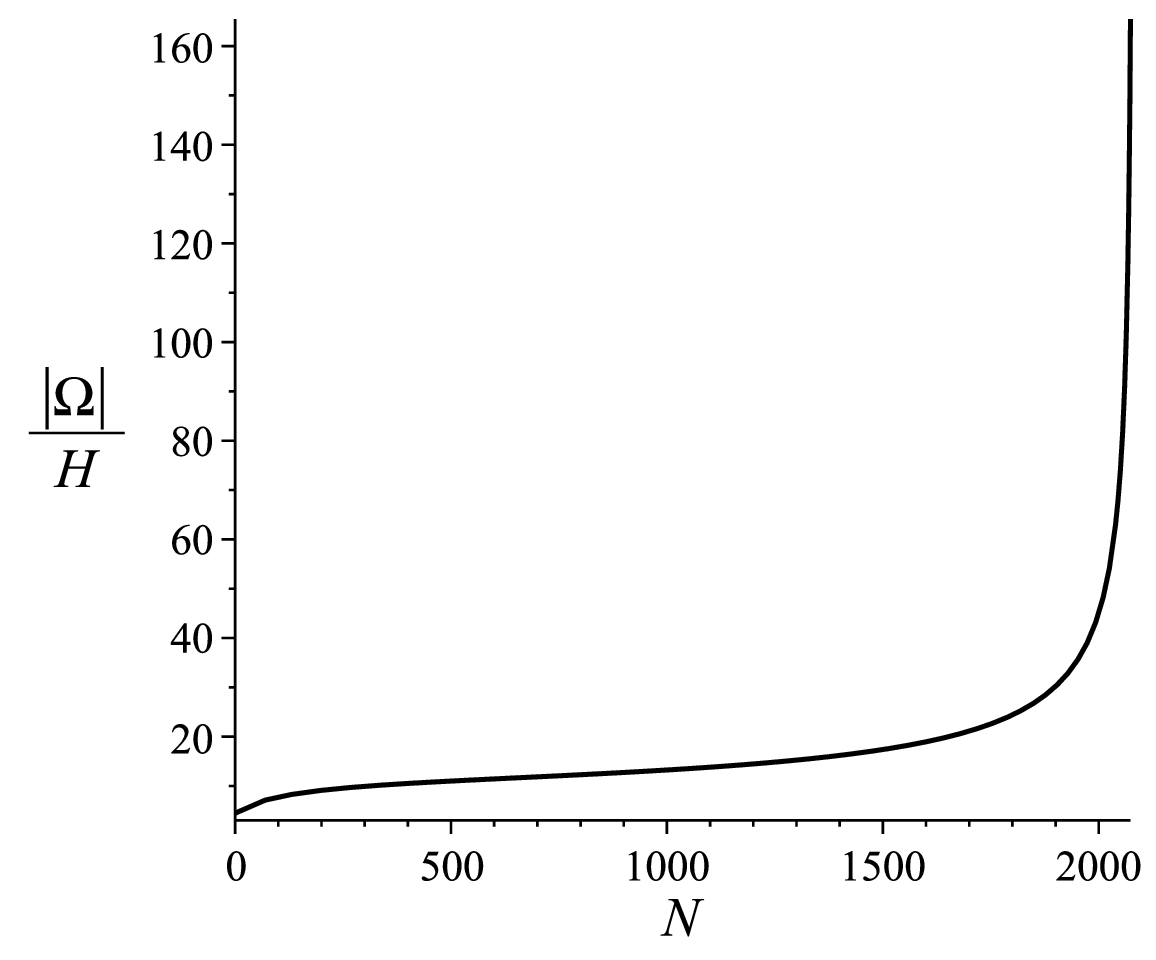}
\hspace*{0.7cm}
\includegraphics[scale=0.3385]{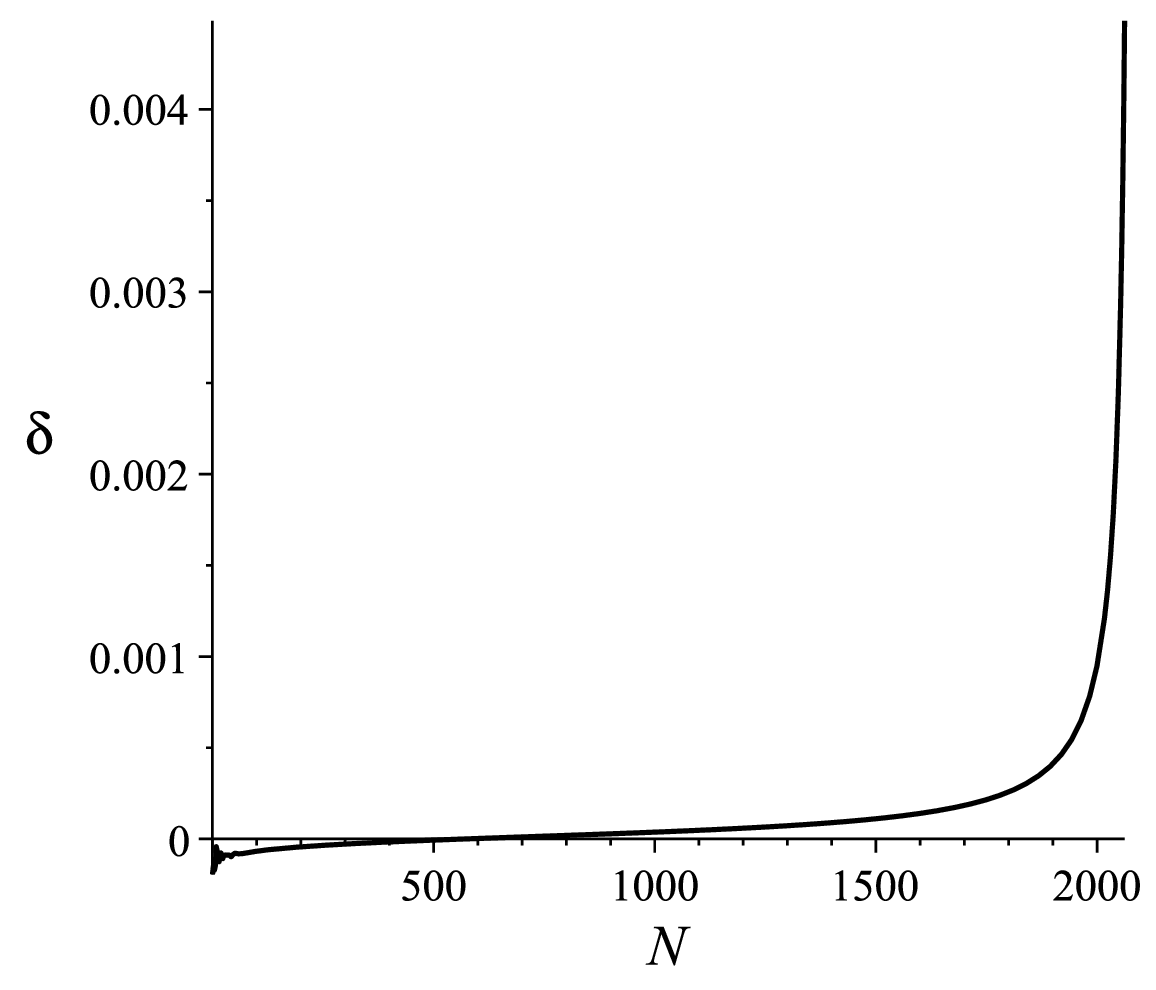}
\end{center}
\vspace{-0.7cm}
\caption{{\small On the left: Plot of $|\eta_{\perp}| = |\Omega| / H$ determined by (\ref{eta_perp_simeq}), for the example (\ref{U_NI})-(\ref{Par_values}). On the right: Plot of the function $\delta$ defined in (\ref{def_delta}), for the same example.}}
\label{Eta_perp_abs}
\vspace{0.1cm}
\end{figure}
Here, we will illustrate numerically (\ref{Th_const}), which is also an implicit check on (\ref{CC_full}) as is clear from Section \ref{SecCharAng}. For that purpose, let us take the following form of $U$:
\be \label{U_NI}
U(\phi) = 1 + \cos \!\left( \frac{\phi}{f} \right) \quad ,
\ee
where $f = const$\,; this is the potential in natural inflation. In line with \cite{GSRPR}, we also take the following values for the model parameters:
\be \label{Par_values}
M = 10^{-3} \qquad , \qquad m_h = 10 \qquad , \qquad f = 10 \quad .
\ee
Note that, with the help of (\ref{dPhdN}), one can convert every function of $\phi$ into a function of the number of e-folds $N$. In the example given by (\ref{U_NI})-(\ref{Par_values}), the slow-roll approximation is well-satisfied for about $2000$ e-folds, while $\varepsilon \sim {\cal O} (1)$ is reached at around $N \sim 2100$\,.

\begin{figure}[t]
\begin{center}
\hspace*{-0.5cm}
\includegraphics[scale=0.45]{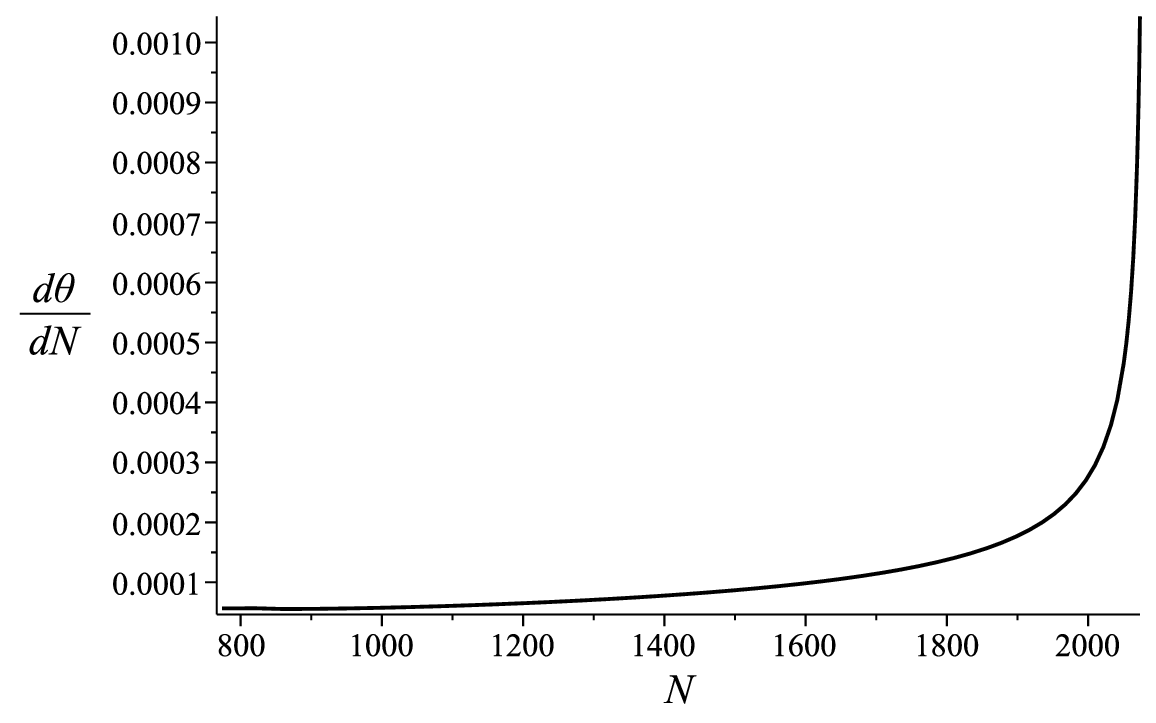}
\end{center}
\vspace{-0.7cm}
\caption{{\small Plot of $\frac{d\vartheta}{dN}$\,, determined by (\ref{cos_f}), for the example (\ref{U_NI})-(\ref{Par_values}).}}
\label{dThdN}
\vspace{0.1cm}
\end{figure}
In Figure \ref{Eta_perp_abs}, we illustrate the dependence of the turning rate on $N$ and also, as a preliminary check on our considerations, perform a comparison with \cite{GSRPR}. More precisely, on the left of the Figure we have plotted $|\eta_{\perp}|$\,, determined by (\ref{eta_perp_simeq}) together with (\ref{dPhdN}), for the example (\ref{U_NI})-(\ref{Par_values}). On the right of Figure \ref{Eta_perp_abs}, we have plotted, for the same example, the function: 
\be \label{def_delta}
\delta \equiv \frac{|\eta_{\perp}|-|\eta_{\perp}^*|}{|\eta_{\perp}|} \,\,\,\, ,
\ee
where $\eta_{\perp}^*$ is the dimensionless turning rate obtained from the expression \cite{GSRPR}:
\be \label{Et_*}
\eta_{\perp}^* \simeq \frac{m_h}{H} \frac{\frac{\sqrt{2}\chi}{M}}{\left(1+\frac{2\chi^2}{M^2} \right)^{1/2}} \,{\rm sign}(U') \,\,\,\, .
\ee
Clearly, (\ref{eta_perp_simeq}) agrees numerically with (\ref{Et_*}) to order $10^{-3}$, or even (much) better, during the entire inflationary period. Finally, in Figure \ref{dThdN} we plot $\frac{d \vartheta}{dN}$\,, determined by (\ref{cos_f}) together with (\ref{dPhdN}). Obviously, the numerical agreement with (\ref{Th_const}) is excellent. Of course, the derivation of (\ref{cos_f}) neglected multiple terms containing the slow-roll parameters. Such terms, however, clearly cannot change the conclusion that the magnitude of $\frac{d \vartheta}{dN}$ is very small during the rapid-turn and slow-roll inflationary regime.

\subsection{Angular inflation} \label{AngInfl}

Angular inflation, found in \cite{CRS}, provides another prominent example of a rapid turn model. In this case, the field-space metric and potential can be written in the form:
\be \label{Ang_rth_GV}
ds^2_G \,= \,6 \alpha \,\frac{(dr^2 + r^2 d\tilde{\theta}^2)}{\left( 1 - r^2 \right)^2} \qquad , \qquad V \,= \,\frac{\alpha}{6} \,m^2 \,r^2 \,( \cos^2 \!\tilde{\theta} + R \sin^2 \!\tilde{\theta} ) \quad ,
\ee
where \,$\alpha, m, R = const$ \,and the field $\tilde{\theta}$ should not be confused with the characteristic angle $\vartheta$ of the previous sections. Angular inflation occurs for the following parameter ranges:
\be \label{Ang_par_constr}
\alpha \ll 1 \qquad {\rm and} \qquad R \gtrsim 10 \quad,
\ee
while its field-space trajectory is given by \cite{CRS}:
\be \label{Ang_traj}
r^2 \simeq 1 - \frac{9}{2} \,\frac{\alpha \,( \cot \tilde{\theta} + R \tan \tilde{\theta} )^2}{(R-1)^2} \quad .
\ee

Using (\ref{Ang_rth_GV}), one can compute the Hessian components $V_{nn}$\,, $V_{n \tau}$ and $V_{\tau \tau}$ in this model, just as in the side-tracked example of the previous subsection. Then, substituting (\ref{Ang_traj}), one obtains those components as functions of $\tilde{\theta}$ and the model parameters. In view of (\ref{Ang_par_constr}), it is convenient (and enough) for our present purposes to consider only the leading terms of these expressions, in the limit of small $\alpha$ and large $R$\,. So, expanding in $\alpha <\!\!< 1$ and $R >\!\!> 1$\,, we find:
\bea \label{Vxx_ang}
V_{nn} &\simeq& - \,\frac{1}{2} \,m^2 \,\alpha \,R \,\sin^2 \tilde{\theta} \,\tan^2 \tilde{\theta} \, + \, ... \,\,\, ,\nn \\
V_{n \tau} &\simeq& - \,\frac{1}{2} \,m^2 \,\alpha \,R \,\sin^2 \tilde{\theta} \,\tan \tilde{\theta} \, + \, ... \,\,\, , \nn \\
V_{\tau \tau} &\simeq& \frac{1}{2} \,m^2 \,\alpha \,R \,\sin^2 \tilde{\theta} \,\tan^2 \tilde{\theta} \, + \, ... \,\,\, ,
\eea
where the subleading ``$...$" terms, at each order in $\alpha$\,, contain a series of decreasing powers of $R$\,. More precisely, at ${\cal O} (\alpha)$ we have subleading terms with $R^0$, $R^{-1}$, $R^{-2}$ etc., while at ${\cal O} (\alpha^2)$ and higher we have terms with $R$, $R^0$, $R^{-1}$ etc. Analogously, one can obtain:
\be \label{Vang_l}
V \, \simeq \, \frac{1}{6} \,m^2 \,\alpha \,R \,\sin^2 \tilde{\theta} \, + ... \,\,\, ,
\ee
with a similar structure of the subleading ``$...$" terms. 

From (\ref{Vxx_ang}) we can see that, to leading order in $\alpha <\!\!< 1$ and $R >\!\!> 1$\,, relation (\ref{Vnn_m_Vtt}) is satisfied in this model. In addition, one can verify that (\ref{E2_apr}) is satisfied too, in accordance with our discussion in Subsection \ref{Vnn_eq_m_Vtt}. Indeed, by using (\ref{Vxx_ang})-(\ref{Vang_l}), we obtain, to leading order, the same for both sides of (\ref{E2_apr}):
\be
3 V V_{nn}^2 \,\simeq \,V_{n \tau}^2 V_{\tau \tau} \, \simeq \, \frac{1}{8} \,m^6 \,\alpha^3 \,R^3 \,\sin^6 \tilde{\theta} \,\tan^4 \tilde{\theta} \, + \, {\rm subleading} \,\,\, ,
\ee
in agreement with \cite{AL2}.\footnote{Note that the multitude of subleading terms in $3 V V_{nn}^2$ does not coincide with those in $V_{n \tau}^2 V_{\tau \tau}$\,.} Despite that, condition (\ref{Cons_Ineq}) is problematic now. Indeed, from (\ref{Vxx_ang}) we have that to leading order:
\be
\frac{V_{\tau \tau} |V_{nn}|}{V_{n \tau}^2} \, \simeq \, \tan^2 \tilde{\theta} \,\,\, .
\ee
However, at large \,$\tan^2 \tilde{\theta}$ \,relation (\ref{Ang_traj}) breaks down \cite{CRS}. So the considerations in our Subsection \ref{Vnn_eq_m_Vtt} are essential for understanding the consistency of angular inflation, along the field-space trajectory (\ref{Ang_traj}).

Note that substituting (\ref{Vxx_ang}) in (\ref{eta_perp_simeq}) gives to leading order:
\be
\eta_{\perp}^2 \, \simeq \, 9 \,\tan^2 \tilde{\theta} \,\,\, .
\ee
Hence, in view of the discussion above, the magnitude of $\eta_{\perp}^2$ cannot be very large in this model. On the other hand, the conditions (\ref{RT_w}) are easily satisfied. Indeed, let us verify, for instance, that $\eta_{\perp}^2 >\!\!> \varepsilon$\,. Using the equation of motion $\frac{1}{2} \dot{\sigma}^2 = - \dot{H}$ together with (\ref{c_def})\,, as well as that $3H^2 \cong V$ during slow roll, we obtain:
\be \label{ep_gen}
\varepsilon \, = \, - \frac{\dot{H}}{H^2} = \frac{\dot{\sigma}^2}{2 H^2} = \frac{c^2 V_I V^I}{2 H^4} \cong \frac{9 c^2 V_I V^I}{2 V^2} \,\,\,\, .
\ee
Now, substituting (\ref{c_sol}) in (\ref{ep_gen}) gives:
\be \label{ep_V}
\varepsilon \, \simeq \, \frac{V_I V^I V_{n \tau}^2}{2 V^2 (V_{nn}^2 + V_{n \tau}^2)} \,\,\,\, .
\ee
Note that, to leading order in small $\alpha$ and large $R$\,, one has:
\be
V_I V^I \, \simeq \, \frac{3}{8} \,m^4 \,\alpha^3 \,R^2 \,\sin^2 \tilde{\theta} \,\tan^4 \tilde{\theta} \,\,\,\, .
\ee
Using this and (\ref{Vxx_ang})-(\ref{Vang_l}) in (\ref{ep_V}), we finally find:
\be \label{ep_V_ang}
\varepsilon \, \simeq \, \frac{27}{4} \,\alpha \,\tan^2 \tilde{\theta} \,\,\,\, .
\ee
So, clearly, the condition $\alpha <\!\!< 1$ ensures that both $\varepsilon <\!\!< 1$ and $\eta_{\perp}^2 >\!\!> \varepsilon$\,. And, of course, (\ref{ep_V_ang}) affirms that $\tan^2 \tilde{\theta}$ \,should not be very large, in line with the above discussion.

\section{Conclusions}

In this paper, we reconsidered the issue of the consistency of rapid-turn inflation. Although slow-roll inflationary models with multiple scalars have been studied for a long time (see, for instance, \cite{GWBM,NvT,PT,PT2}), the realization that in the multifield context one does not need a slow-turn approximation is rather recent. This rapid-turn regime leads to novel effects and opens new avenues for embedding the model into a fundamental framework, as we have pointed out in the introduction. However, satisfying the equations of motion, {\it together with all relevant approximations} that determine the slow-roll and rapid-turn regime, is rather nontrivial. The consistency conditions studied here relate the field space metric to the scalar potential. Thus they enable one to determine whether a particular model could allow for the desired inflationary regime, without solving the equations of motion. Alternatively, they enable one to determine in what parts of field space or parameter space, for a given pair of scalar field-space metric and scalar potential, there could be inflationary trajectories that are slow-rolling and rapid-turning. Hence, these conditions facilitate the search for inflationary solutions of that type by scanning large sets of models; in that regard, see \cite{WIA}. 

The most important result of this paper is that we have resolved a potential problem regarding the sustainability of the rapid-turn regime, which was raised recently. Namely, it was claimed that the evolution of the characteristic angle $\vartheta$, which relates the kinematic field-space basis to the potential-gradient one, should lead to a fast exit from the regime of slow-roll and rapid-turn inflation \cite{AL}. Here we studied the behavior of $\vartheta$ by taking into account a specific rapid-turn relation between this angle and the components of the Hessian of the potential, instead of relying on generic arguments about those components. Then, by using the generalized consistency condition (\ref{CC_full}), we showed that $\vartheta$ remains (nearly-)constant, as a function of the number of e-folds, throughout the entire slow-roll and rapid-turn period. This proves that this inflationary regime can be consistently sustained for a large number of e-folds.

\section*{Acknowledgements}

\noindent I would like to thank A. Achucarro, A. Buchel, P. Christodoulidis, O. Iarygina and C. Lazaroiu for useful discussions or correspondence regarding cosmological inflation and, specifically, the rapid-turn regime. I have received partial support from the Bulgarian NSF grant KP-06-N68/3.

\appendix

\section{Technical details regarding $\vartheta$-equation} \label{Deriv_Eq_Th}
\setcounter{equation}{0}

This Appendix traces the steps in the derivation of (\ref{EoMTh_ph_2}) in a streamlined manner, while introducing some notation that will be convenient for the present purposes.

As preparation, note that from the definitions (\ref{N_def}) and (\ref{BasesRel}) one has:
\be \label{phi_dot_sig}
\dot{\varphi}^I = \dot{\sigma} \,(n^I \cos \vartheta + \tau ^I \sin \vartheta) \,\,\, .
\ee
In addition, the definitions (\ref{ndef}) imply the following relations:
\be \label{nabla_nt_rels}
n^J \nabla_J n^I = \frac{V_{n \tau}}{\hat{N}} \,\tau^I \,\,\, , \,\,\,\, \tau^J \nabla_J n^I = \frac{V_{\tau \tau}}{\hat{N}} \,\tau^I \,\,\, , \,\,\,\, n^J \nabla_J \tau^I = - \frac{V_{n \tau}}{\hat{N}} \,n^I \,\,\, , \,\,\,\, \tau^J \nabla_J \tau^I = - \frac{V_{\tau \tau}}{\hat{N}} \,n^I \,\, ,
\ee
where for convenience we have denoted:
\be \label{Nhat_V}
\hat{N} \equiv \sqrt{V_I V^I} \,\,\, .
\ee
Indeed, to understand, for instance, the second relation in (\ref{nabla_nt_rels}), consider:
\be
\tau_I \tau^J \nabla_J n^I = \tau_I \tau^J \nabla_J \!\left( \frac{V^I}{\sqrt{V_K V^K}} \right) = \tau_I \tau^J \frac{\nabla_J V^I}{\sqrt{V_K V^K}} + \tau_I \tau^J V^I \nabla_J \!\left( \frac{1}{\sqrt{V_K V^K}} \right) = \frac{V_{\tau \tau}}{\sqrt{V_K V^K}} \,\, ,
\ee
where we have used that $\tau_I V^I = 0$ according to (\ref{ndef}). Similarly, one can verify all relations in (\ref{nabla_nt_rels}) by using solely the definitions in (\ref{ndef}).

Now, projecting (\ref{EoM_sc}) along $\tau_I$ gives:
\be \label{EoM_sc_t_proj}
\tau_I D_t \dot{\varphi}^I + 3 H \dot{\sigma} \sin \vartheta = 0 \,\,\, ,
\ee
due to (\ref{phi_dot_sig}) and (\ref{ndef}). Then, using (\ref{phi_dot_sig})-(\ref{nabla_nt_rels}), one computes:
\be \label{Eq_t_proj}
\tau_I D_t \dot{\varphi}^I \equiv \tau_I \dot{\varphi}^J \nabla_J \dot{\varphi}^I = \ddot{\sigma} \sin \vartheta + \dot{\sigma} \dot{\vartheta} \cos \vartheta + \dot{\sigma}^2 \cos \vartheta \left( \frac{V_{n \tau}}{\hat{N}} \cos \vartheta + \frac{V_{\tau \tau}}{\hat{N}} \sin \vartheta \right) \,.
\ee
Substituting (\ref{Eq_t_proj}) in (\ref{EoM_sc_t_proj}) leads to the equation:
\be \label{Eq_th_proj_t}
\dot{\vartheta} + (3 - \eta_{\parallel}) H \tan \vartheta + \frac{\dot{\sigma}}{\hat{N}} \left( V_{n \tau} \cos \vartheta + V_{\tau \tau} \sin \vartheta \right) \,= \,0 \,\,\,\, ,
\ee
where we have also used the definition of $\eta_{\parallel}$ in (\ref{SR_par}).

Similarly, projecting (\ref{EoM_sc}) along $n_I$ and using (\ref{phi_dot_sig})-(\ref{nabla_nt_rels}), one obtains:
\be \label{Eq_th_proj_n}
\dot{\vartheta} - (3 - \eta_{\parallel}) H \cot \vartheta + \frac{\dot{\sigma}}{\hat{N}} \left( V_{n \tau} \cos \vartheta + V_{\tau \tau} \sin \vartheta \right) - \frac{\hat{N}}{\dot{\sigma} \sin \vartheta} \,= \,0 \,\,\,\, .
\ee
Note that, from (\ref{c_def}) and (\ref{eta_cos_sin_c}), one has:
\be \label{s_dot_N_hat}
\dot{\sigma} = \frac{\hat{N} \cos \vartheta}{(\eta_{\parallel} -3) H} \,\,\, . 
\ee
This implies that (\ref{Eq_th_proj_n}) coincides with (\ref{Eq_th_proj_t}). Hence there is, indeed, a single equation which determines $\vartheta (t)$, as should be the case.

Let us now write this $\vartheta$-equation in a more convenient form. Substituting (\ref{s_dot_N_hat}) in (\ref{Eq_th_proj_t}) gives:
\be
\dot{\vartheta} + (3 - \eta_{\parallel}) H \tan \vartheta - \frac{\cos \vartheta}{(3-\eta_{\parallel}) H} \left( V_{n \tau} \cos \vartheta + V_{\tau \tau} \sin \vartheta \right) \,= \,0 \,\,\,\, ,
\ee
which can be written as:
\be \label{dTh_dN}
\frac{d\vartheta}{dN} + (3 - \eta_{\parallel}) \tan \vartheta - \frac{\cos \vartheta}{(3-\eta_{\parallel}) H^2} \left( V_{n \tau} \cos \vartheta + V_{\tau \tau} \sin \vartheta \right) \,= \,0 \,\,\,\, ,
\ee
in terms of the number of e-folds $N$ defined as usual by $dN \!\equiv \!H dt$\,.\footnote{The number of e-folds $N$ should not be confused with the notation $\hat{N}$\,, introduced in (\ref{Nhat_V}).}
Finally, in the slow-roll approximation, in which $3H^2 \cong V$\,, the $\vartheta$-equation (\ref{dTh_dN}) becomes:
\be
\frac{d\vartheta}{dN} + 3 \tan \vartheta - \frac{\cos \vartheta}{V} \left( V_{n \tau} \cos \vartheta + V_{\tau \tau} \sin \vartheta \right) \,\cong \,0 \,\,\,\, .
\ee


\begin{thebibliography}{100}

\bibitem{OOSV}
G. Obied, H. Ooguri, L. Spodyneiko and C. Vafa, {\em De Sitter Space and the Swampland}, arxiv:1806.08362 [hep-th].

\bibitem{GK}
S. Garg and C. Krishnan, {\em Bounds on Slow Roll and the de Sitter Swampland}, JHEP 11 (2019) 075,  arXiv:1807.05193 [hep-th].

\bibitem{OPSV}
H. Ooguri, E. Palti, G. Shiu, C. Vafa, {\em Distance and de Sitter Conjectures on the Swampland}, Phys. Lett. B 788 (2019) 180, arXiv:1810.05506 [hep-th].

\bibitem{AP}
A. Achucarro and G. Palma, {\em The string swampland constraints require multi-field inflation}, JCAP 02 (2019) 041, arXiv:1807.04390 [hep-th].

\bibitem{BPR}
R. Bravo, G. A. Palma, S. Riquelme, {\em A Tip for Landscape Riders: Multi-Field Inflation Can Fulfill the Swampland Distance Conjecture}, JCAP 02 (2020) 004, 1906.05772 [hep-th].

\bibitem{PSZ} G. A. Palma, S. Sypsas, C. Zenteno, {\em Seeding primordial black holes in multi-field inflation}, Phys. Rev. Lett. 125 (2020) 121301, arXiv:2004.06106 [astro-ph.CO].

\bibitem{FRPRW} J. Fumagalli, S. Renaux-Petel, J. W. Ronayne, L. T. Witkowski, {\em Turning in the landscape:~a new mechanism for generating Primordial Black Holes}, Phys. Lett. B 841 (2023) 137921, arXiv:2004.08369 [hep-th].

\bibitem{LA}
L. Anguelova, {\em On Primordial Black Holes from Rapid Turns in Two-field Models}, JCAP 06 (2021) 004, arXiv:2012.03705 [hep-th].

\bibitem{LA2_pbh} L. Anguelova, {\em Primordial Black Hole Generation in a 
Two-field Inflationary Model}, Springer Proc. Math. Stat. 396 (2022) 193,  
arXiv:2112.07614 [hep-th].

\bibitem{AB} A. Brown, {\em Hyperinflation}, Phys. Rev. Lett. 121 (2018)
251601, arXiv:1705.03023 [hep-th].

\bibitem{SM}
S. Mizuno, S. Mukohyama, {\em Primordial perturbations from inflation with a hyperbolic field-space}, Phys. Rev. D 96 (2017) 103533, arXiv:1707.05125 [hep-th].

\bibitem{CRS} P. Christodoulidis, D. Roest, E. Sfakianakis, {\em Angular inflation in multi-field $\alpha$-attractors}, JCAP 11 (2019) 002, arXiv:1803.09841 [hep-th].

\bibitem{GSRPR} S. Garcia-Saenz, S. Renaux-Petel, J. Ronayne, {\em Primordial fluctuations and non-Gaussianities in sidetracked inflation}, JCAP 1807 (2018) 057, arXiv:1804.11279 [astro-ph.CO].

\bibitem{ACIPWW} A. Achucarro, E. Copeland, O. Iarygina, G. Palma, D.G. Wang, Y. Welling, {\em Shift-Symmetric Orbital Inflation:~single field or multi-field?}, Phys. Rev. D 102 (2020) 021302, arXiv:1901.03657 [astro-ph.CO].

\bibitem{TB} T. Bjorkmo, {\em The rapid-turn inflationary attractor}, Phys. Rev. Lett. 122 (2019) 251301, arXiv:1902.10529 [hep-th].

\bibitem{BM} T. Bjorkmo, M. C. D. Marsh, {\em Hyperinflation generalised:  from its attractor mechanism to its tension with the `swampland conditions'}, JHEP 04 (2019) 172, arXiv:1901.08603 [hep-th].

\bibitem{ChRoSf}
P. Christodoulidis, D. Roest, E. Sfakianakis, {\em Scaling attractors in multi-field inflation}, JCAP 12 (2019) 059, arXiv:1903.06116 [hep-th].

\bibitem{APR2}
V. Aragam, S. Paban, R. Rosati, {\em Multi-field Inflation in High-Slope Potentials}, JCAP 04 (2020) 022, arXiv:1905.07495 [hep-th].

\bibitem{APR}
V. Aragam, S. Paban, R. Rosati, {\em The Multi-Field, Rapid-Turn Inflationary Solution}, JHEP 03 (2021) 009, arXiv:2010.15933 [hep-th].

\bibitem{CR}
P Christodoulidis, R. Rosati, {\em (Slow-)Twisting inflationary attractors}, JCAP 09 (2023) 034, arXiv:2210.14900 [hep-th].

\bibitem{IMS}
O. Iarygina, M.C.D. Marsh, G. Salinas, {\em Non-Gaussianity in rapid-turn multi-field inflation}, JCAP 03 (2024) 014, arXiv:2303.14156 [astro-ph.CO].

\bibitem{ChG}
P. Christodoulidis, J.-O. Gong, {\em Enhanced power spectra from multi-field inflation}, arXiv:2311.04090 [hep-th].

\bibitem{ASSV}
Y. Akrami, M. Sasaki, A. Solomon, V. Vardanyan, {\em Multi-field dark energy: cosmic acceleration on a steep potential}, Phys. Lett. B819 (2021) 136427, 	arXiv:2008.13660 [astro-ph.CO].

\bibitem{ADGW}
L. Anguelova, J. Dumancic, R. Gass, L.C.R. Wijewardhana, {\em Dark Energy from Inspiraling in Field Space}, JCAP 03 (2022) 018, arXiv:2111.12136 [hep-th]

\bibitem{EASV}
J. Eskilt, Y. Akrami, A. Solomon, V. Vardanyan, {\em Cosmological dynamics of multifield dark energy}, Phys. Rev. D 106 (2022) 023512, arXiv:2201.08841 [astro-ph.CO].

\bibitem{ADGW2}
L. Anguelova, J. Dumancic, R. Gass, L.C.R. Wijewardhana, {\em Dynamics of Inspiraling Dark Energy}, Eur. Phys. J. C 84 (2024) 365, arXiv:2311.07839 [hep-th].

\bibitem{BFM}
T. Bjorkmo, R. Z. Ferreira, M.C.D. Marsh, {\em Mild Non-Gaussianities under
Perturbative Control from Rapid-Turn Inflation Models}, JCAP 12 (2019) 036,
arXiv:1908.11316 [hep-th].

\bibitem{CAP}
S. Cespedes, V. Atal, G. A. Palma, {\em On the importance of heavy fields during inflation}, JCAP 05 (2012) 008, arXiv:1201.4848 [hep-th].

\bibitem{EGNO}
J. Ellis, M. Garcia, D. Nanopoulos, K. Olive, {\em A No-Scale Inflationary Model to Fit Them All}, JCAP 08 (2014) 044, arXiv:1405.0271 [hep-ph].

\bibitem{ACPRZ} V. Aragam, R. Chiovoloni, S. Paban, R. Rosati,
I. Zavala, {\em Rapid-turn inflation in supergravity is rare and
tachyonic}, JCAP 03 (2022) 002, arXiv:2110.05516 [hep-th].

\bibitem{AL}
L. Anguelova, C.I. Lazaroiu, {\em Dynamical consistency conditions for rapid turn inflation}, JCAP 05 (2023) 020, arXiv:2210.00031 [hep-th].

\bibitem{AL2}
L. Anguelova, C.I. Lazaroiu, {\em Consistency Condition for Slow-roll and Rapid-turn Inflation}, arXiv:2311.18683 [hep-th].

\bibitem{AAGP}
A. Achucarro, V. Atal, C. Germani, G. A. Palma, {\em Cumulative effects in inflation with ultra-light entropy modes}, JCAP 02 (2017) 013, arXiv:1607.08609 [astro-ph.CO].

\bibitem{HP} A. Hetz, G. Palma, {\em Sound Speed of Primordial
Fluctuations in Supergravity Inflation}, Phys. Rev. Lett. 117 (2016) 101301, arXiv:1601.05457 [hep-th].

\bibitem{AGHPP} A. Achucarro, J.-O. Gong, S. Hardeman, G. Palma,
S. Patil, {\em Features of heavy physics in the CMB power spectrum}, JCAP 01 (2011) 030, arXiv:1010.3693 [hep-ph].

\bibitem{CCLBNZ} D. Chakraborty, R. Chiovoloni, O. Loaiza-Brito,
G. Niz, I. Zavala, {\em Fat inflatons, large turns and the $\eta$-problem}, JCAP 01 (2020) 020, arXiv:1908.09797 [hep-th].

\bibitem{WIA}
R. Wolters, O. Iarygina, A. Achucarro, {\em Generalised conditions for rapid-turn inflation}, arXiv:2405.11628 [astro-ph.CO].

\bibitem{ChRS}
P. Christodoulidis, D. Roest, E. I. Sfakianakis, {\em Attractors, Bifurcations and Curvature in Multi-field Inflation}, JCAP 08 (2020) 006, arXiv:1903.03513 [gr-qc].



\bibitem{GWBM}
C. Gordon, D. Wands, B. A. Bassett, R. Maartens, {\em Adiabatic and entropy
perturbations from inflation}, Phys. Rev. D 63 (2001) 023506, astro-ph/0009131.

\bibitem{NvT}
S. G. Nibbelink, B. van Tent, {\em Scalar perturbations during multiple field slow-roll inflation}, Class. Quant. Grav. 19 (2002) 613-640, hep-ph/0107272.

\bibitem{PT}
C. M. Peterson, M. Tegmark, {\em Testing Two-Field Inflation}, Phys. Rev. D 83 (2011) 023522, arXiv:1005.4056 [astro-ph.CO].

\bibitem{PT2}
C. M. Peterson, M. Tegmark, {\em Non-Gaussianity in Two-Field Inflation}, Phys. Rev. D 84 (2011) 023520, arXiv:1011.6675 [astro-ph.CO].

\end{thebibliography}
\end{document}